\documentclass[aps,prl,twocolumn]{revtex4-2}
\usepackage{CJK}
\usepackage{amsmath}

  \usepackage[dvips]{graphicx}

\begin{document}


\title{Independent Component Analysis for Signal Crosstalk Elimination in Infrared Interferometry.}


\author{L.~Esteban}
\email[]{lesteban@nebrija.es}
\affiliation{ARIES Research Center, Universidad Antonio de Nebrija, Madrid, Spain.}
\author{A.~Regadio}
\affiliation{Instituto Nacional de T\'ecnica Aeroespacial (INTA), Madrid, Spain.}
\author{U.~Losada}
\affiliation{Centro de Investigaciones Energ\'eticas, Medioambientales y Tecnol\'ogicas (CIEMAT), Madrid, Spain.}


\date{\today}

\begin{abstract}

Infrared interferometers are optical devices that can measure optical path-length differences by measuring changes in the refractive index. Several arrangements can be deployed, from single channel devices to  multichannel double color heterodyne interferometers. These type of devices are typically used to recover the spatial electron density profile of fusion plasmas. This process involves measuring precisely the phase differences of the intermediate frequencies. A source of error that affects the measurements is the crosstalk that appears due to the coupling of the signals into the different channels of the interferometers. The inter-channel coupling of the signals is extremely difficult to eliminate specially when the probing frequencies are close to each other. In this paper it is shown that Independent Component Analysis effectively eliminates inter-crosstalk coupling in such devices. Furthermore, it is shown how the Signal-to-Noise ratio is dramatically increased when this technique is used.

\end{abstract}

\pacs{}

\maketitle


%

\textit{Introduction.-}A key parameter in experimental thermonuclear reactors is the line integrated electron density which is typically measured, for high densities using Infrared (IR) interferometers. For a non-relativistic plasma ($T_e<20\;\mathrm{keV}$) a beam crossing the plasma would suffer a phase shift proportional to the line integrated density given by:

\begin{equation}
\phi=\mathrm{r_e}\lambda\int_{\mathrm{plasma}}n_e(z)\mathrm{d}z,
\label{eq1}
\end{equation}

\noindent where $\mathrm{r_e}$ is the classical electron radius, $\lambda$ is the wavelength of the probing beam and $n_e(z)$ is the electron density. Therefore, by measuring this phase shift the line-integrated electron density can be recovered. Another beam with the same geometrical length as the probing one that does not cross the plasma is used as a reference. Both beams interfere in a square law detector. In the case of Infrared (IR) heterodyne interferometers the optical frequency is shifted by means of an Acousto Optical Modulators (AOMs). Therefore, the detected interference signal in the detector will have a frequency coincident with the one driving the AOM and in an ideal scenario a phase shift proportional to the line integrated electron density.

The typical wavelengths used for these type of devices range from $1\;\mu\mathrm{m}$ to $10\;\mu\mathrm{m}$. With such low wavelengths mechanical vibrations become an important issue and a complementary interferometer sharing the same optical path as the first one is necessary to cancel these vibrations. This second interferometer will be less sensitive to plasma density fluctuations. In this context, the line integrated electron density can be computed according to the following expression \cite{01Hutchinson2002, 02Esteban2010}:

\begin{equation}\label{LID}
\int{n_e\mathrm{d}l=}\frac{\Delta\phi_{\mathrm{1}}\lambda_{1}-\Delta\phi_{2}\lambda_{2}}{r_e(\lambda_1^2-\lambda_2^2)},
\end{equation}

\noindent where $\Delta\phi_{\mathrm{1}}$ is the phase difference between the interference signal of the first interferometer and its correspondent reference. It will contain information about mechanical vibrations, thermal drifts and plasma density variations. $\Delta\phi_{\mathrm{2}}$ is the phase difference between the signals of the second interferometer, it mainly contains information about mechanical vibrations and thermal drifts. Thus, by subtracting both optical paths $\Delta\phi_{\mathrm{1}}\lambda_{1}-\Delta\phi_{2}\lambda_{2}$ and  provided that the sources of error are minimized \cite{03Esteban2011, 04Esteban2013} the line electron integrated density can be computed.

\begin{figure}[!ht]
\centering
\includegraphics[width=8.8cm]{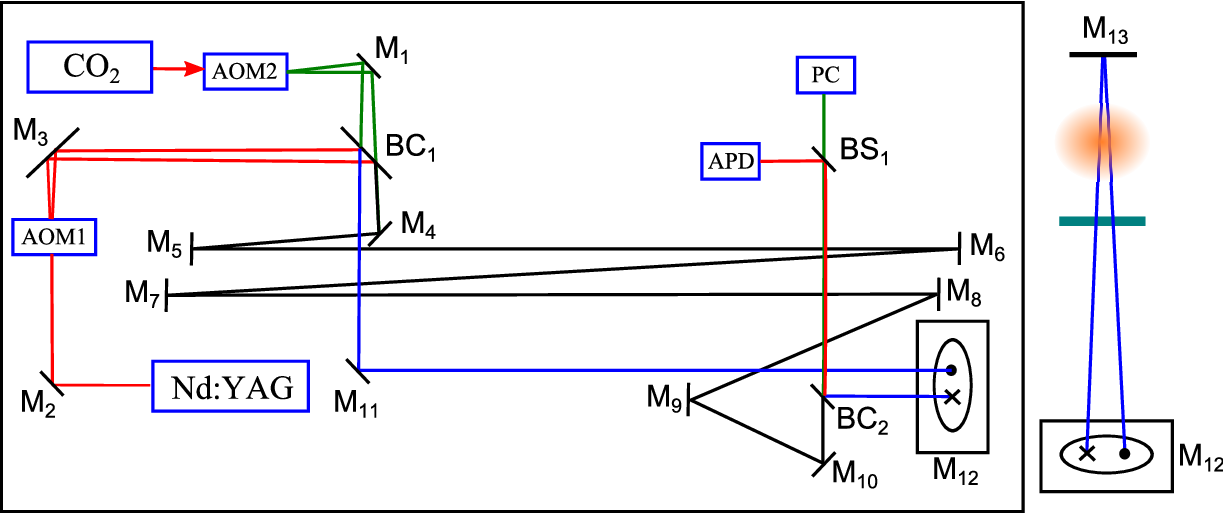}
\caption{TJ-II IR interferometer layout.}
\label{layout}
\end{figure}

Typically two detectors are used, one to the detect the first wavelength $\lambda_1$ and the second for the other one $\lambda_2$. The Layout of the TJ-II IR interferometer is shown on Figure \ref{layout} taken from \cite{05Esteban2012}. This is a double color $\mathrm{CO_2}-10.591\mu\mathrm{m}$ $\mathrm{Nd:YAG}-1.064\mu\mathrm{m}$ heterodyne interferometer where the modulation frequencies of the AOMs are in the order of $40\;\mathrm{MHz}$ for both wavelengths.

\textit{Crosstalk in IR interferometers.-}Because in double color arrangements two interferometers share the same optical paths it is necessary to separate the beams by means of  beam splitters. Since the setup is not perfect both interferometers can affect each other causing optical crosstalk. Also, if analog intermediate frequency mixing stages are used (in the back-end electronics) crosstalk of electrical origin can also appear \cite{06Sanchez2008}.



In an interferometer without crosstalk the output interference signals are given by:

\begin{eqnarray}\label{hola2}
  y_1 &= \sin\left(\omega_1\cdot t+\Delta\phi_1\right), \\
  y_2 &= \sin\left(\omega_2\cdot t+\Delta\phi_2\right),
\end{eqnarray}

\noindent where $\omega_1$ and $\omega_2$ are the angular heterodyne frequencies of both interferometers, usually their values are close to each other. $\Delta\phi_1$ and $\Delta\phi_2$ are the phase displacements that the probing beams of both interferometers suffer when crossing the plasma due  mainly to plasma density fluctuations and mechanical vibrations. In the presence of crosstalk, instead of having two clean signals a mixture of $y_1$ and $y_2$ appears in both outputs of the detectors, $s_1$ and $s_2$ as:

\begin{equation}\label{matrices}
\begin{bmatrix}
s_{1} \\
s_{2}
\end{bmatrix}=
\begin{bmatrix}
\alpha_{11} & \alpha_{12} \\
\alpha_{21} & \alpha_{22}
\end{bmatrix}\cdot
\begin{bmatrix}
y_{1} \\
y_{2}
\end{bmatrix}
\end{equation}

\noindent the terms $\alpha_{ij}$ are the coupling coefficients. Using a more compact notation:
\begin{equation}
\label{mixing}
\mathrm{\textbf{S}}=\mathrm{\textbf{A}}\cdot\mathrm{\textbf{Y}}
\end{equation}
\noindent where $\mathrm{\textbf{A}}$ is the mixing matrix, $\mathrm{\textbf{S}}$ is a vector containing the measured signals and $\mathrm{\textbf{Y}}$ contains the independent components:

Several techniques have been employed to obtain the clean signals $\mathrm{\textbf{Y}}$ from the mixtures $\mathrm{\textbf{S}}$. In \cite{06Sanchez2008} an iterative algorithm to remove the crosstalk is used. This algorithm uses a gradient descent based approach, it subtracts a weighted and shifted sum of the aggressor signal from the victim and minimizes the Interference-to-Signal power Ratio (ISR) of the line integrated electron density Eq. \ref{LID} by adjusting a coupling coefficient. This was a first approach to cope with the crosstalk in the TJ-II interferometer that effectively attenuated it in some cases, but and since the ISR is not a convex function the solution obtained sometimes falls into local minima.

Other approaches try use heterodyne frequencies that are separated enough so they can be split electronically. For instance, in a test-bench that was deployed for the W7-X stellarator \cite{02Esteban2010, 07Esteban2011} two separated heterodyne frequencies were used $40\;\mathrm{MHz}$ for the first wavelength $\mathrm{CO_2}$ and $25\;\mathrm{MHz}$ for the second one, CO. The two signals were separated using digital filtering in a Field Programmable Gate Array (FPGA). This type of solution is highly dependable on the optical setup and is not always achievable.

In \cite{08Allen1999, 09Hernandez2013} a Fast Fourier Transform (FFT) based crosstalk removal algorithm for the LIGO gravitational interferometer \cite{10Abramovici1996, 11Barish2018, 12Abbott2016, 13Abbott2017} is presented. This technique removes the crosstalk in the frequency domain. First the N-point FFT of the measured signals is computed and after the crosstalk coefficient is obtained by projecting the spectra of the signals over each other for a set of split intervals. In the TJ-II IR interferometer this same technique was optimized and implemented in an FPGA for its operation in real-time in the TJ-II IR interferometer \cite{09Hernandez2013}.

In this paper a technique based in Independent Component Analysis to remove the crosstalk in IR interferometers is proposed. Moreover, it is also shown how this technique can be used together with digital filtering to split interference signals in systems that use a single detectors for several wavelengths. Not only the signals are split but also their gain is automatically adjusted.


\textit{Independent Component Analysis.-} Recalling the crosstalk removal problem of Eq. \ref{matrices} it resembles the cocktail party problem in audio signal processing \cite{14Shlens2014}, where out of a mixture of sounds in the environment one or more sources must be separated. This type of filtering problem is called Blind Source Separation (BSS) because no information about the non-corrupted signals is given. Independent Component Analysis (ICA) \cite{15Hyvarinen2000} is a widely used BSS algorithm that solves this problem not only in the audio separation context but in many other applications including the separation of biomedical signals \cite{16James2004}, image processing or feature extraction \cite{17Gonzalez2009}. In this sense, as above mentioned the crosstalk removal problem in IR interferometers is a similar problem as the cocktail party one and thus it can be removed using ICA.


\textit{ICA theoretical background.-} In Eq. \ref{mixing} by finding the inverse of matrix $\mathrm{\textbf{A}}$, the original signals can be recovered:

\begin{equation}\label{mixing2}
    \mathrm{\textbf{Y}}=\mathrm{\textbf{A}}^{-1}\cdot\mathrm{\textbf{S}}=\mathrm{\textbf{W}}\cdot\mathrm{\textbf{S}},
\end{equation}
\noindent where matrix $\mathrm{\textbf{A}}^{-1}$ is also known as the weight matrix $\mathrm{\textbf{W}}$.

\begin{figure}[!ht]
\begin{center}
\includegraphics[width=4cm]{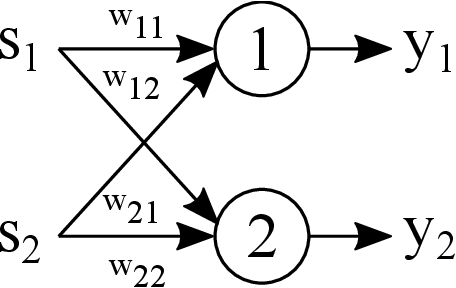}
\caption{ICA equivalent Neural Network.}
\label{neuronas}
\end{center}
\end{figure}

The process of finding the non-contaminated signals, $y_1$ and $y_2$ could be seen as the process of training or finding the weights of the two neuron network of figure \ref{neuronas}.

As above mentioned the training process to find the weight matrix can be done using ICA. ICA tries to find the uncorrelated non-gaussian sources from their mixtures, in other words finds the weight matrix \textbf{W}.

Two are the most used ICA algorithms that allow to obtain the un-mixing matrix \textbf{W}, projection pursuit and fastICA \cite{18Tharwat2021}.

In this contribution fastICA is used to remove the crosstalk in IR interferometers. In order to do so an estimation of the gaussianity of the outputs, $y_1$ and $y_2$ of the neural network on Figure \ref{neuronas} has to be made. Since the not contaminated signals are uncorrelated non-gaussian, the weight matrix \textbf{W} can be found by reducing the gaussianity of $y_1$ and $y_2$, in other words of \textbf{Y}.

In particular, in this work the fastICA algorithm is used. FastICA uses an approximation to negative entropy (negentropy) to maximize the non-gaussianity using a fixed-point iteration scheme \cite{15Hyvarinen2000}:

\begin{equation}\label{negentropy}
   J\left(y\right)\simeq \sum_{i=1}^p k_i \left[E\left(G_i(y)\right)-E\left(G_i(\nu)\right)\right]^2,
\end{equation}

\noindent where $J$ is the negentropy, $\nu$ is a random variable with a standard normal distribution, $k_i$ are positive constants and $G_i$ are non-quadratic functions. For instance:

\begin{eqnarray}
  G_1\left(u\right)=\frac{1}{a\cdot s_1}\log\cosh a_1 u, \\
  G_2\left(u\right)=-\mathrm{e}^{\left(\frac{-u^2}{2}\right)}.
\end{eqnarray}

\noindent where $a$ is a constant, $1 \leq a \leq 2$.


\textit{ICA pre-processing.-}Before applying the fastICA algorithm some preprocessing of the data has to be made:

\begin{enumerate}
  \item The mean of the input signals must be substracted:
  \begin{equation}\label{matrices2}
\begin{bmatrix}
r_{1} \\
r_{2}
\end{bmatrix}=
\begin{bmatrix}
s_{1}-\mu \\
s_{2}-\mu
\end{bmatrix}
\end{equation}

\noindent where $\mu$ is the mean of all mixture signals, in this case $s_1$ and $s_2$. The mean can be added back once the independent components are computed by the ICA algorithm.

  \item The centered data has to be whitened, the signals must be decorrelated. To do so an Eigen-Value-Decomposition (EVD) of the covariance matrix of the centered data \textbf{R} has to be carried out:

\begin{equation}\label{covariancem}
    \sum=E\left(\mathrm{\textbf{R}}\mathrm{\textbf{R}}^T\right),
\end{equation}

\noindent where \textbf{R} is the matrix containing the centered data. The EVD decomposition is given by:

\begin{equation}\label{EVD}
    \sum=\mathrm{\textbf{U}}^{-1}\mathrm{\Lambda}\mathrm{\textbf{U}},
\end{equation}

\noindent U is the orthogonal matrix of eigenvectors and $\Lambda$ is the diagonal matrix of eigenvalues. Finally, the centered data can be obtained as follows:
\begin{equation}\label{EVD2}
    \mathrm{\textbf{C}}=\mathrm{\textbf{U}}\mathrm{\textbf{R}},
\end{equation}

      \item In order to further process the data using ICA the decorrelated signals must also have unit variance. Therefore, they must be scaled:

  \begin{equation}\label{EVD3}
    \mathrm{\textbf{B}}=\sqrt{\Lambda}\mathrm{\textbf{C}},
\end{equation}

\noindent and $\sqrt{\Lambda}=\left[\sqrt{\lambda_1}, \sqrt{\lambda_2}, \cdots, \sqrt{\lambda_n}\right]$, where the terms $\lambda_i$ are the eigenvalues.

\end{enumerate}


\textit{fastICA algorithm.-} Once the signals have been centered, whitened and scaled they can be furthered processed using the fastICA algorithm. As it has been already mentioned the fastICA algorithm finds independent components by maximizing their non-gaussianity. To do so the negentropy is used. The fastICA algorithm proceeds as it is described in \cite{15Hyvarinen2000} and summarized below, if matrix \textbf{B} contains the centered whitened signals:

\begin{enumerate}
  \item Choose an initial random weight vector \textbf{W}.
  \item Update the weight vector as:
    \begin{equation}\label{wupdate}
    \mathrm{\textbf{W}}^+=E\left(\mathrm{\textbf{B}}g\left(\mathrm{\textbf{W}}^T \mathrm{\textbf{B}}\right)\right)-E\left(g'\left(\mathrm{\textbf{W}}^T \mathrm{\textbf{B}}\right)\right)\mathrm{\textbf{W}},
\end{equation}
  \item normalize the weight vector:
  \begin{equation}\label{wnorm}
   \mathrm{\textbf{W}}=\frac{\mathrm{\textbf{W}}^+}{\|\mathrm{\textbf{W}}^+\|}
\end{equation}

  \item Repeat the process until the weights converge.

\end{enumerate}


\textit{Results.-}The fastICA algorithm has been applied to experimental data obtained from the TJ-II IR interferometer. The fastICA algorithm is applied to the output interference signals from a photoconductor detector for the $\mathrm{CO_2}$ wavelength and an Avalanche Photo Detector (APD) for the $\mathrm{NdYAG}$ one Figure \ref{layout}. In this measurement an analog intermediate frequency stage that downconverts the baseband signals from $\sim 40\;\mathrm{MHz}$ to $\sim 1\;\mathrm{MHz}$ is used. The down-converted signals are sampled with a $12\;\mathrm{bits}$ ADC at a rate of $8\;\mathrm{MSPS}$.

\begin{figure}[!ht]
\centering
\includegraphics[width=8.8cm]{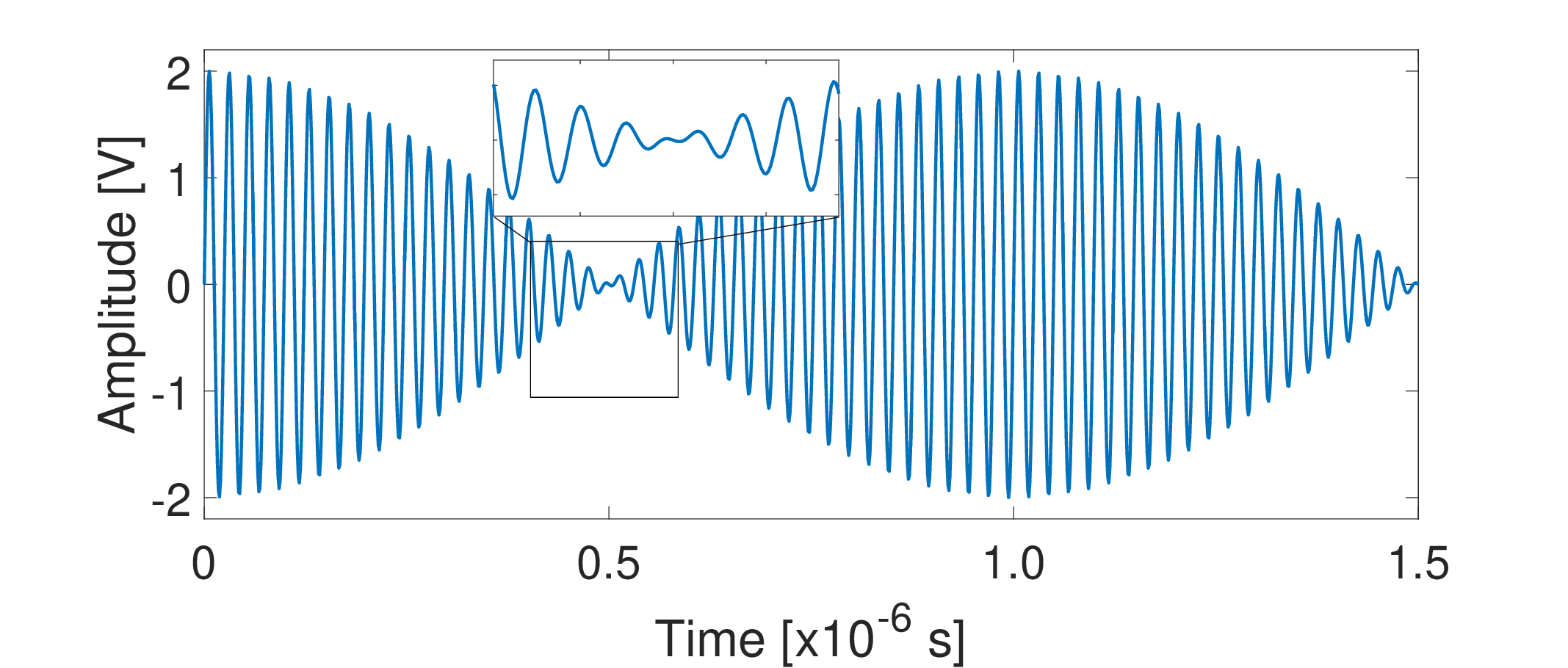}
\caption{Strong coupling of both interferences signals.}
\label{modulacion}
\end{figure}

\begin{figure}[!ht]
\centering
\includegraphics[width=8.8cm]{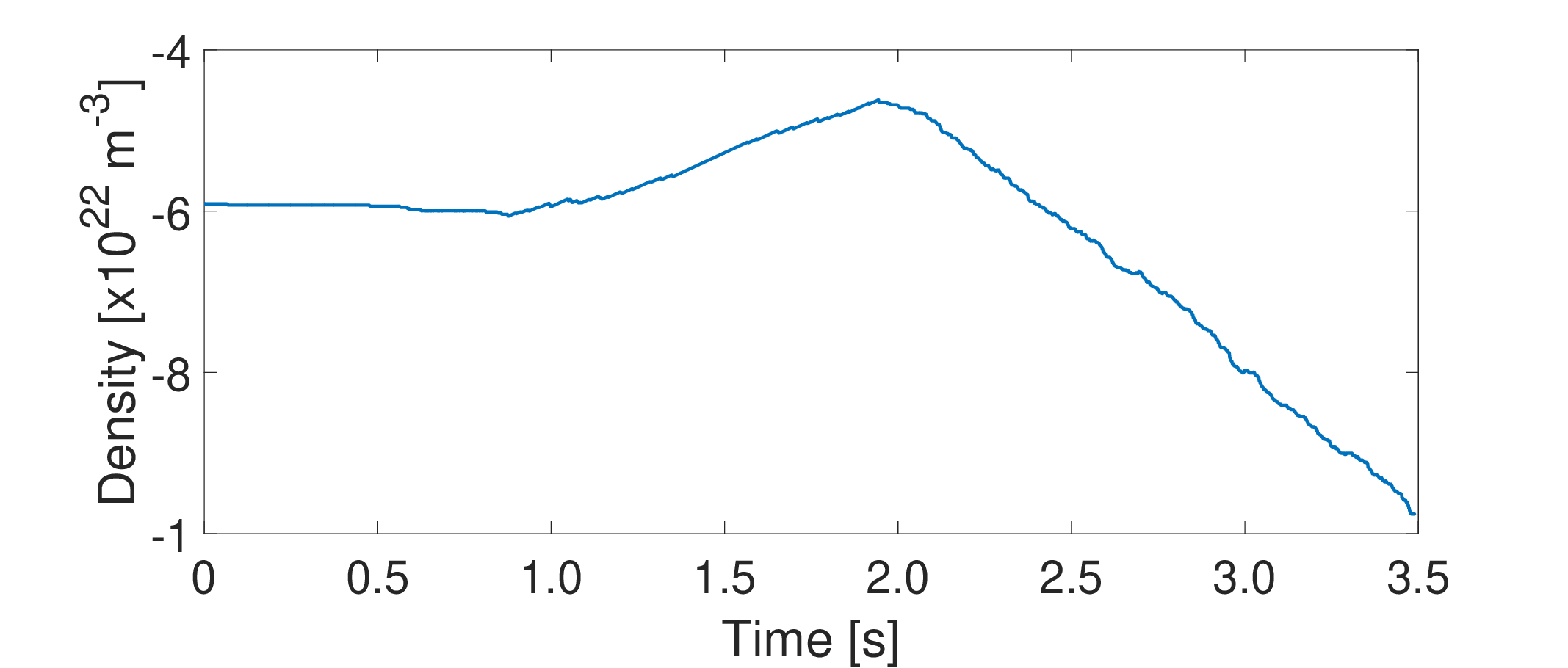}
\caption{TJ-II electron density signal from the IR interferometer without ICA based crosstalk correction.}
\label{crossu}
\end{figure}

These data is then centered and whitened following the procedure explained above. Once this is done the fastICA algorithm is applied to the data. The data analyzed in this paper is strongly affected by crosstalk which appears in the form of an envelope over the output signal as it can be seen in Figure \ref{modulacion}. On the presence of crosstalk two effects can be observed:

\begin{enumerate}
  \item Degradation of the SNR in the final electron density \cite{06Sanchez2008}.

  \item The algorithm cannot measure the phases on zero crossings of the envelope signal. Thus, making impossible the recovery of the electron density profile, Figure \ref{crossu}.
\end{enumerate}

\begin{figure}[!ht]
\centering
\includegraphics[width=9cm]{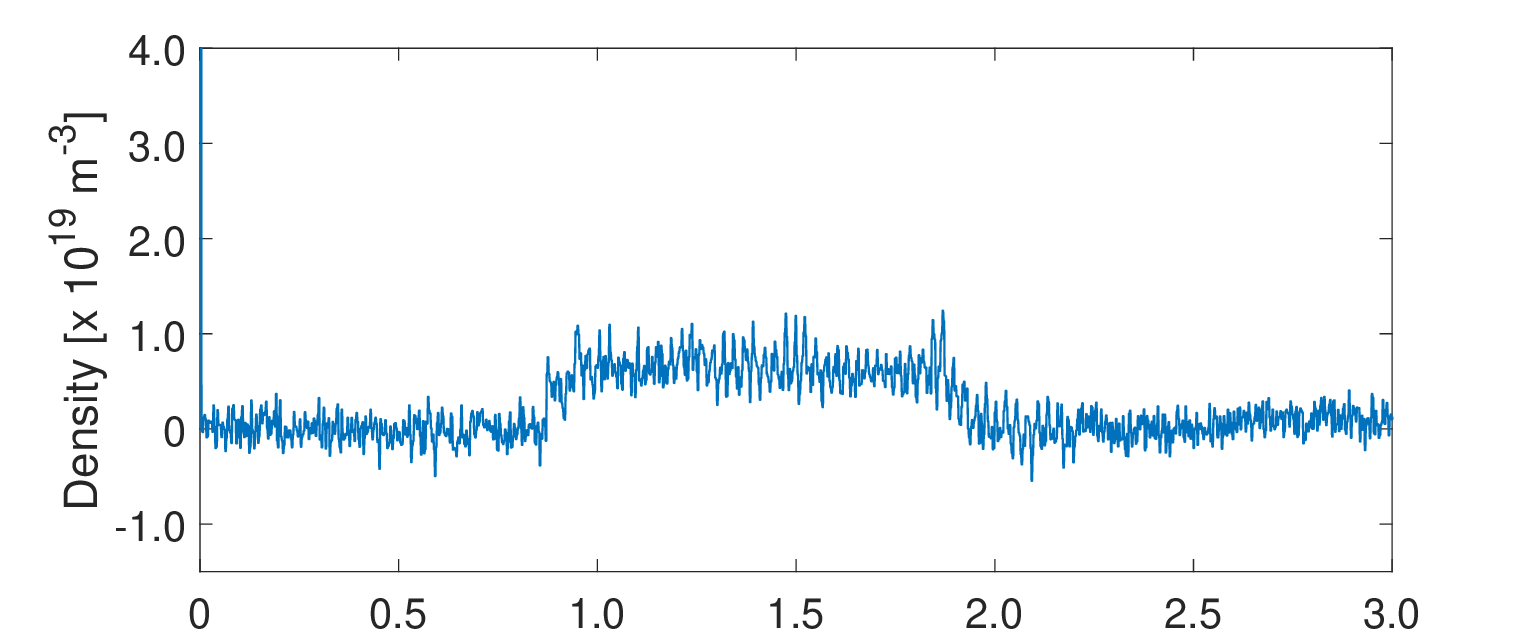}
\caption{TJ-II electron density signal from the IR interferometer with ICA based crosstalk correction.}
\label{crossc}
\end{figure}

In Figure \ref{crossu} it is shown how the line electron density cannot be measured under the given levels of crosstalk. Namely due to the fact that the phase measuring algorithm cannot keep track of the zero crossing in the nulls of the envelope signal. On the other hand if the fastICA algorithm is applied to the interference signals the independent components are recovered using matrix $\mathrm{\textbf{W}}$ allowing the computation of the electron density as it shown in Figure \ref{crossc}. This effect has been observed in several discharges in the TJ-II IR interferometer \cite{03Esteban2011}.


So far a configuration that uses two detectors, one for each wavelength has been described. However, another approach is to use a single detector to detect both colors or wavelengths \cite{02Esteban2010, 19Innocente1997}. Therefore, the signals must be split electronically. A good approach is to use bandpass filters. In \cite{02Esteban2010} Finite Impulse Response (FIR) are implemented digitally for this task. To achieve a good Interference to Signal Ratio ISR high order filters are required, typically $p>29$. In this paper it is shown that digital filtering can be enhanced by using ICA to both relax the order $p$ of the filters and to achieve an almost perfect rejection of the coupled signals.

\begin{figure}[!ht]
\centering
\includegraphics[width=8.4cm]{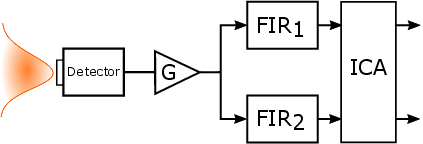}
\caption{Frequency diplexion scheme based in FIR filtering and ICA signal decomposition.}
\label{esquema}
\end{figure}

\begin{figure}[!ht]
\centering
\includegraphics[width=8.8cm]{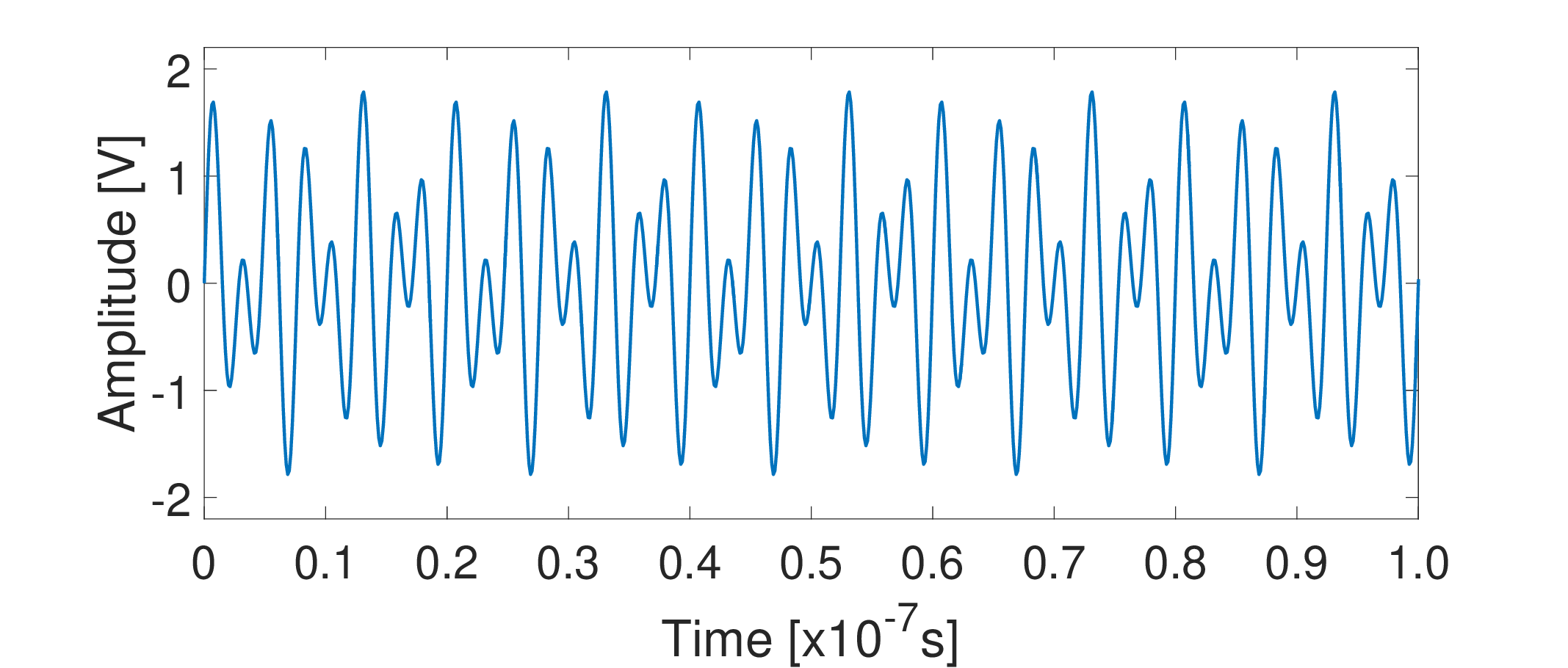}
\caption{Signal to be split containing $25\;\mathrm{MHz}$ and $40\;\mathrm{MHz}$ frequency tones.}
\label{mezcla}
\end{figure}

\begin{figure}[!ht]
\centering
\includegraphics[width=8.8cm]{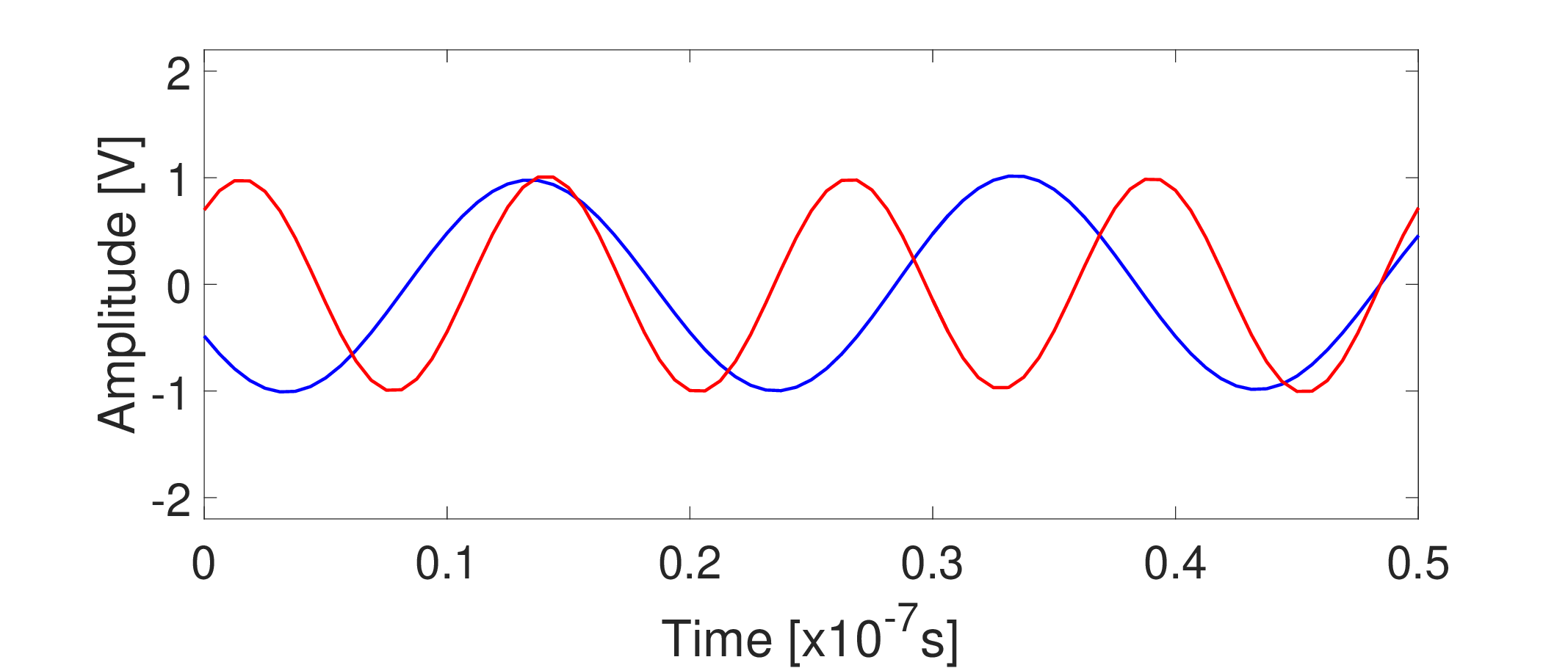}
\caption{$25\;\mathrm{MHz}$ and $40\;\mathrm{MHz}$ split signals.}
\label{separadas}
\end{figure}

The architecture of the detection system is shown in Figure \ref{esquema}. The FIR filters used are 5th order bandpass filter that use a Hamming window. The input to the filters is the signal shown on Figure \ref{mezcla} that is a mixture of two tones $25\;\mathrm{MHz}$ and $40\;\mathrm{MHz}$. At the outputs of the filters two signals are obtained each one of them having different weights that multiply the independent components (interference signals). Since the filters used are bandpass the mean of the signals is automatically removed. These signals are then fed into the fastICA framework to first whiten the data and finally to obtain the diplexed signals as shown in Figure \ref{separadas}. As it can be seen in the Figure the signals are perfectly split with zero mean and amplitude one.

\textit{Conclusions.-} The use of infrared interferometry for the line electron integrated measurement requires the implementation of highly precise phase meters. Therefore, it is essential to minimize systematic error sources, and to evaluate globally the errors made in the measurements. A limiting error is the crosstalk that can be of optical or electrical origin and it can degrade the input signals to the point that the line integrated electron density cannot be recovered. A strait-forward solution for one channel interferometers was proposed in \cite{06Sanchez2008}. However for double color arrangements or specially for multichannel interferometers the problem is more complicated and the matrix $\mathrm{\textbf{W}}$ that relates each of the measured signals with the independent signals must be found. The ideal signals can be determined by multiplying the corrupted ones by matrix $\mathrm{\textbf{W}}$. In this paper it has been shown that this can be done by using ICA and in particular the fastICA algorithm. The methodology has been proved using signals from the TJ-II IR interferometer affected strongly by crosstalk. It has been shown how by obtaining the independent components using the fastICA algorithm the measured signals can be corrected and the final electron density calculated. In addition in the paper a technique based on FIR filters and ICA is proposed for systems using single detectors for several wavelenghts. This approach allows to both relax the order $p$ of the filters used and to achieve an almost perfect rejection of the coupled signals. The methodology described in this paper has been applied to IR interferometers for nuclear fusion reactors but it can be applied to other type of interferometers such as LIGO \cite{10Abramovici1996} or the future LISA \cite{20Schwarze2019} due to the fact that crosstalk removal is of central importance for gravitational wave detection.

\bibliographystyle{unsrt}
\bibliography{bib}

\begin{thebibliography}{10}

\bibitem{01Hutchinson2002}
Ian~H Hutchinson.
\newblock Principles of plasma diagnostics.
\newblock {\em Plasma Physics and Controlled Fusion}, 44(12):2603--2603, 2002.

\bibitem{02Esteban2010}
L~Esteban, Miguel S{\'a}nchez, J~S{\'a}nchez, Petra Kornejew, Matthias Hirsch,
  JA~L{\'o}pez, A~Fern{\'a}ndez, and O~Nieto-Taladriz.
\newblock Continuous phase measurement in the w7-x infrared interferometer by
  means of a fpga and high-speed adcs.
\newblock {\em Fusion Science and Technology}, 58(3):771--777, 2010.

\bibitem{03Esteban2011}
Luis~Esteban Hern{\'a}ndez.
\newblock {\em High precission fpga based phase meters for infrared
  interferometer fusion diagnostics}.
\newblock PhD thesis, Universidad Polit{\'e}cnica de Madrid, 2011.

\bibitem{04Esteban2013}
Luis Esteban, JA~L{\'o}pez, E~Sedano, S~Hernandez-Montero, and Miriam Sanchez.
\newblock Quantization analysis of the infrared interferometer of the tj-ii
  stellarator for its optimized fpga-based implementation.
\newblock {\em IEEE Transactions on Nuclear Science}, 60(5):3592--3596, 2013.

\bibitem{05Esteban2012}
Luis Esteban and Miguel S{\'a}nchez.
\newblock Interferometry for fusion devices.
\newblock In {\em Interferometry: Research and Applications in Science and
  Technology}. INTECH, 2012.

\bibitem{06Sanchez2008}
M~S{\'a}nchez, L~Esteban, P~Kornejew, and M~Hirsch.
\newblock Admissible crosstalk limits in a two colour interferometers for
  plasma density diagnostics. a reduction algorithm.
\newblock In {\em AIP conference Proceedings}, volume 993, pages 187--190.
  American Institute of Physics, 2008.

\bibitem{07Esteban2011}
Luis Esteban, Miguel S{\'a}nchez, Juan~Antonio L{\'o}pez, Petra Kornejew,
  Matthias Hirsch, and Octavio Nieto-Taladriz.
\newblock Development of efficient fpga-based multi-channel phase meters for
  ir-interferometers.
\newblock {\em IEEE Transactions on Nuclear Science}, 58(4):1562--1569, 2011.

\bibitem{08Allen1999}
Bruce Allen, Wensheng Hua, and Adrian Ottewill.
\newblock Automatic cross-talk removal from multi-channel data.
\newblock {\em arXiv preprint gr-qc/9909083}, 1999.

\bibitem{09Hernandez2013}
S~Hern{\'a}ndez-Montero, JA~L{\'o}pez, M~S{\'a}nchez, L~Esteban, and
  CA~L{\'o}pez.
\newblock Real time fpga-based crosstalk elimination for multichannel
  interferometry systems in fusion diagnostics.
\newblock {\em IEEE Transactions on Nuclear Science}, 60(5):3585--3591, 2013.

\bibitem{10Abramovici1996}
A~Abramovici, W~Althouse, J~Camp, D~Durance, JA~Giaime, A~Gillespie,
  S~Kawamura, A~Kuhnert, T~Lyons, FJ~Raab, et~al.
\newblock Improved sensitivity in a gravitational wave interferometer and
  implications for ligo.
\newblock {\em Physics Letters A}, 218(3-6):157--163, 1996.

\bibitem{11Barish2018}
Barry~C Barish.
\newblock Nobel lecture: Ligo and gravitational waves ii.
\newblock {\em Reviews of Modern Physics}, 90(4):040502, 2018.

\bibitem{12Abbott2016}
Benjamin~P Abbott, Richard Abbott, TDe Abbott, MR~Abernathy, Fausto Acernese,
  Kendall Ackley, Carl Adams, Thomas Adams, Paolo Addesso, RX~Adhikari, et~al.
\newblock Observation of gravitational waves from a binary black hole merger.
\newblock {\em Physical review letters}, 116(6):061102, 2016.

\bibitem{13Abbott2017}
Benjamin~P Abbott, Richard Abbott, TD~Abbott, F~Acernese, K~Ackley, C~Adams,
  T~Adams, P~Addesso, Rana~X Adhikari, Vaishali~B Adya, et~al.
\newblock Gw170814: a three-detector observation of gravitational waves from a
  binary black hole coalescence.
\newblock {\em Physical review letters}, 119(14):141101, 2017.

\bibitem{14Shlens2014}
Jonathon Shlens.
\newblock A tutorial on independent component analysis.
\newblock {\em arXiv preprint arXiv:1404.2986}, 2014.

\bibitem{15Hyvarinen2000}
Aapo Hyv{\"a}rinen and Erkki Oja.
\newblock Independent component analysis: algorithms and applications.
\newblock {\em Neural networks}, 13(4-5):411--430, 2000.

\bibitem{16James2004}
Christopher~J James and Christian~W Hesse.
\newblock Independent component analysis for biomedical signals.
\newblock {\em Physiological measurement}, 26(1):R15, 2004.

\bibitem{17Gonzalez2009}
Rafael~C Gonzalez.
\newblock {\em Digital image processing}.
\newblock Pearson Education, 2009.

\bibitem{18Tharwat2021}
Alaa Tharwat.
\newblock Independent component analysis: An introduction.
\newblock {\em Applied Computing and Informatics}, 17(2):222--249, 2021.

\bibitem{19Innocente1997}
P~Innocente, S~Martini, A~Canton, and L~Tasinato.
\newblock Upgrade of the rfx co2 interferometer using in-vessel optics for
  extended edge resolution.
\newblock {\em Review of scientific instruments}, 68(1):694--697, 1997.

\bibitem{20Schwarze2019}
Thomas~S Schwarze, Germ{\'a}n~Fern{\'a}ndez Barranco, Daniel Penkert, Marina
  Kaufer, Oliver Gerberding, and Gerhard Heinzel.
\newblock Picometer-stable hexagonal optical bench to verify lisa phase
  extraction linearity and precision.
\newblock {\em Physical review letters}, 122(8):081104, 2019.

\end{thebibliography}
\end{document}